\documentclass[sigconf]{acmart}

\usepackage[toc,page]{appendix}

\usepackage{placeins}
\usepackage{array}
\usepackage{pdfpages}

\usepackage{amsmath}
\usepackage{booktabs}
\usepackage{textcomp}

\usepackage{graphicx}
\usepackage{float}
\usepackage[caption = false]{subfig}
\usepackage{caption}

\usepackage{pgfplots}
\usepackage{changepage}
\pgfplotsset{ticks=none}
\pgfplotsset{compat=1.14}
\pgfplotsset{width=4cm,compat=1.9}

\usepackage[justification=centering]{caption}
\renewcommand\footnotetextcopyrightpermission[1]{}
\begin{document}

\title{Evolutionary Deep Learning to Identify Galaxies in the Zone of Avoidance}

\author{David Jones$^{1}$, Anja Schroeder$^{2}$, Geoff Nitschke$^{1}$}
\affiliation{
  \institution{$^1$Department of Computer Science}
  \institution{University of Cape Town}
  \institution{$^2$South African Astronomical Observatory}
    \country{South Africa}
}
\email{jnsdav026@myuct.ac.za, anja@hartrao.ac.za, gnitschke@cs.uct.ac.za}

\begin{abstract}
The \textit{Zone of Avoidance} makes it difficult for astronomers to catalogue galaxies at low latitudes to our galactic plane due to high star densities and extinction. However, having a complete sky map of galaxies is important in a number of fields of research in astronomy. There are many unclassified sources of light in the Zone of Avoidance and it is therefore important that there exists an accurate automated system to identify and classify galaxies in this region. This study aims to evaluate the efficiency and accuracy of using an evolutionary algorithm to evolve the topology and configuration of \textit{Convolutional Neural Network} (CNNs) to automatically identify galaxies in the Zone of Avoidance. A supervised learning method is used with data containing near-infrared images. Input image resolution and number of near-infrared passbands needed by the evolutionary algorithm is also analyzed while the accuracy of the best evolved CNN is compared to other CNN variants.
\end{abstract}


\maketitle

\section{INTRODUCTION}
Many areas of research that astronomers are currently dealing with are better understood by knowing the distribution of mass in our local universe. Knowing this distribution is achieved by cataloguing as many celestial objects, including galaxies, across as much of the whole sky as possible \cite{lucas, kraan}. However, the \textit{Zone of Avoidance} (ZoA) obstructs astronomers\textquotesingle\ view of extra-galactic space through its high star, dust, and gas density \cite{kraan}. This makes detecting and identifying galaxies in the ZoA very difficult. Due to the nature of near-infrared light which is less affected by \textit{extinction (see \ref{astro})} in the ZoA, \textit{near-infrared} surveys of the ZoA have had previous success in creating images that allow for easier detection of galaxies \cite{amores}.

A trained human eye is still very capable of manually identifying galaxies in the ZoA. However, due to the vast nature of the data collected and time-consuming process of manually performing the task, the need for an accurate automated system to identify galaxies in the ZoA is evident, with previous attempts to build such a system achieving limited success \cite{kraan}. This automated process requires being able to visually classify galaxies and is a complex task, requiring a machine learning method such as a neural network \cite{khalifa, kim, kriz}. \textit{Deep learning neural networks} have previously been used to classify different types of galaxies from one another outside the ZoA with 97\% accuracy, indicating its potential \cite{khalifa}.

\textit{Convolutional Neural Networks} (CNNs) are neural networks with a ``preprocessing" layer before the actual neural network which can be used to extract features in multi-dimensional array input data \cite{simon}. The structure of CNNs makes it ideal for working with images \cite{conv}, which is essentially what classifying galaxies deals with, making it the obvious neural architecture choice.

CNNs and neural networks in general need to have specifically tuned parameters and topologies to yield the best results for their given task. Non-optimal parameters can yield a CNN that works no better than random chance \cite{schmid}. \textit{Evolutionary algorithms} (EAs) are based on biological evolution and aim to find an optimal solution to a problem by evolving a set of potential solutions from generation to generation. An EA can hence be applied to the parameters and topology of a CNN to find the optimal configuration that will be best for classifying galaxies \cite{streich}.
Previous automated galaxy identification attempts have been made \cite{drink, lewis, lahav, khalifa, kim}, but none have been made using specifically an EA on CNNs\textquotesingle\ topologies, and few have even attempted to target the ZoA.

\subsection{Research Objectives}

A program named \textit{Galyxi Vysion}, which uses an EA to modify CNN topologies, was developed as a potential tool for astronomers to identify galaxies in the ZoA. The objectives for this study are:

\begin{itemize}
	\item Accurately identify galaxies in the ZoA.
    \item Deal with ``noise" from other light sources in the ZoA (known as \textit{source confusion} or \textit{confusion noise}) \cite{lucas, gonz, barav}.
  	\item Produce a CNN configuration that is highly specific to the task of classifying and identifying galaxies at given input dimensions, making it worth the processing time over a human adjusted configuration.
  \item Potentially offer an improved work-flow to astronomers in helping identifying galaxies in the ZoA.
\end{itemize}

The hypotheses of this study are that:

\begin{itemize}
  \item A topologically evolved CNN will be able to identify galaxies in the ZoA since CNNs have already proven to be good at image classification, even with noise \cite{conv}, and EAs have also already been proven effective at CNN parameter tuning \cite{miik}.
  \item The EA will provide a significant benefit in finding an optimal CNN configuration since having correctly adjusted CNN parameters that are difficult to tune manually \cite{schmid}.
  \item Using larger input images and all JHK passbands as opposed to one passband will achieve better results since there is more information and classifying galaxies is a complex task, particularly with the added noise of the ZoA \cite{khalifa, kriz}.
\end{itemize}

\section{BACKGROUND}
\subsection{Mapping Galaxies in the Zone of Avoidance} \label{astro}
The ZoA is the area of space obscured by our own galaxy between roughly $|$b$|$ = -10\textdegree\ to $|$b$|$ = 10\textdegree\ from the Galactic Plane \cite{jarreta}. It is in this area that the density of stars and effects of \textit{extinction} are so high that classifying light sources becomes very difficult \cite{kraanb}. Extinction can be described as the scattering and absorption of light caused by dust and gas in space and plays a large factor in obscuring galaxies in the ZoA \cite{amores, butler, lucas}.

In order for astronomers to understand our universe, information on the statistical distributions of different astronomical objects across our sky need to be obtained by mapping these objects in catalogues \cite{kraan}. One very important catalogue in understanding the distribution of mass in our local universe is the catalogue of galaxies \cite{lucas}. Due to the ZoA, obtaining a complete sky-map of galaxies is difficult but necessary and is needed in many fields of research such as the \textit{dipole} in the \textit{Cosmic Microwave Background} (CMB), peculiar velocity of galaxies in our local cluster \cite{giov}, isotropy of our local universe and understanding the \textit{Great Attractor} \cite{ray}. It is therefore important to be able to identify as many galaxies as possible, even in the densest regions of the ZoA.

The data used for this study consists of FITS files \textit{(see section \ref{data})} containing images of the ZoA with labelled light sources. These images are compiled into what are called \textit{surveys} of areas of the sky \cite{saito}. Near-infrared light includes electromagnetic waves around the ZYJHK passbands with wavelengths from about 0.8 to about 4 \textit{microns} (millionths of a meter), which is out of the visible range. Due to the large extinction effect that the ZoA has on light passing through it, light from galaxies in the ZoA is redder. However, \textit{near-infrared\textit} light is less affected by extinction. Using near-infrared surveys has had success in the past at being able to see more clearly through the ZoA, particularly using JHK passbands, while still being good at seeing galaxies \cite{amores}.

Some surveys done on the ZoA using near-infrared passbands include DENIS (Deep Near-Infrared Survey of the Southern Sky, 1995-2001) \cite{jarreta}, 2MASS (Two Micron All-Sky Survey, 1997-2001) \cite{gonz}, UKIDSS Galactic Plane Survey (The United Kingdom Infrared Deep Sky Survey, 2007-2008) \cite{lucas} and VVV (VISTA variables in the Via Lactea, 2011-present) \cite{saito}. 2MASS and DENIS, being the earlier of the four surveys, were still successful in probing the ZoA to see what lies in it, given that the star density was not too high \cite{lucas}. However, UKIDSS and especially VVV are newer surveys with VVV still collecting data while operating at higher resolutions and a greater sensitivity (ability to detect faint objects) than 2MASS and DENIS \cite{saito}. This makes both UKIDSS and particularly VVV good sources to use to gather data on galaxies in the ZoA.

In order to correctly identify galaxies from other sources of light, certain characteristics or properties about galaxies that can be automatically detected should be known. Galaxies usually have redder J-K colours, unlike stars which are usually bluer \cite{amores}. The next defining character of a galaxy is its shape. Galaxies are said to have different morphologies, including, elliptic, spiral, barred spiral, and irregular or peculiar types \cite{abra}. A galaxy\textquotesingle s shape can generally be approximated by an ellipse, giving galaxies certain ellipticities which can be used to identify them from other sources of light \cite{rah}. If viewed from the side, a spiral galaxy will appear flat with a central bulge. If the galaxy is not viewed side-on, one may even be able to see the arms of the galaxy if it is a spiral galaxy. Galaxies can also be identified by their colour distribution in different passbands and general surface brightness profile \cite{kraan}.

One difficulty that needs to be dealt with is that some galaxies are partially obscured by another light-source such as a star that is much brighter than the galaxy. Another problem area are \textit{low surface brightness diffuse galaxies} since they are often so dim they are barely visible at all and can easily get lost with the other light in the ZoA \cite{butler, abra, jarreta, jarretb}. The last potential issue is that planetary nebula and multi-star systems can sometimes be confused as being galaxies, and may turn up as false positives \cite{jarreta}.

\subsection{Artificial Neural Networks (ANNs) and Deep Learning}

Machine learning can be defined as being a subcategory of Artificial Intelligence (AI) that aims to create machines or systems that automatically learn and improve their behavior over time at doing a certain task, given that they can access relevant data \cite{kots}. Approaches to machine learning include \textit{supervised, unsupervised, semi-supervised, and reinforcement} learning \cite{schmid}. Supervised learning uses labelled training data and implies that humans are the program\textquotesingle s teachers \cite{turner}. This is the approach used for this study since Galyxi Vysion is fed images labelled as either \textit{galaxy} or \textit{non-galaxy}.

\textit{Artificial Neural Networks} (ANNs) are machine learning techniques that are inspired by the way neurons and their connections in biological brains work \cite{baj}.

An ANN consists of processing units that simulate neurons, called \textit{nodes}. The structure of an ANN can be represented as having layers of nodes, with each layer having any number of neurons depending on what is needed \cite{ann}. The first layer is known as the input layer and the last layer is called the output layer. In-between there may also be one or more other layers called \textit{hidden} layers \cite{baj}.

Nodes can have any number of connections between each other, with the possibility of more than one input signal coming from different connections into the node, and only one signal coming out of the node \cite{ann}. The way the nodes and their connections have been laid out is known as the \textit{topology} of the network. Each connection has a weight associated with it, which is the value that the signal fed into the connection is multiplied by \cite{baj}.

ANNs can be described as function approximators which enables them to generalize about a given task or problem. This means they can learn to effectively emulate any function that maps an input \textit{$x$} in a problem space to a solution \textit{$y$} \cite{ann}. ANNs are robust as they are noise tolerant, making them good for galaxy classification as data in the ZoA contains a lot of other light sources and galaxies can come in many variations while being viewed at different angles and distances \cite{rah}.

\textit{Deep learning} is a subfield of machine learning that entails learning multileveled structures and hierarchies of data. Deep learning in the context of ANNs means that there is more than one hidden layer in the network \cite{schmid}. Since galaxy classification is a complex task, some level of deep learning is potentially required \cite{khalifa, kriz}.

\subsubsection{Activation Functions} \label{activations}

A node\textquotesingle\ s output signal is determined by its \textit{activation function}. This function can be linear \textit{($f(x) = \alpha x$)}, where the node is activated if the input signal is greater than a threshold value \textit{theta $\theta$} \cite{ann}. Since not all problems can be classified linearly, a continuous activation function can be used which does not have a threshold value \cite{glor}. Due to the potentially complex nature of identifying a galaxy, a continuous activation function is needed \cite{khalifa}. Some common continuous activation functions include ReLU, Leaky-ReLU, ELU, Sigmoid, TanH and SoftPlus \cite{shar} as seen in \textit{appendix \ref{appendactive}}.

\subsubsection{Optimization Algorithms} \label{optimizations}

In order to approximate the appropriate function, an ANN must learn what the correct connection weights are. This works by repetitively applying trial data to the network and comparing the outputs with an expected correct result \cite{glor}. In supervised learning these example solutions are provided by a human \cite{turner}. At first, the ANN is unlikely to produce the correct output and the connection weights must be adjusted \textit{(optimized)} \cite{glor}.

An example of this is \textit{back-propagation} \cite{bash}. Back-propagation takes the difference between the actual output and the expected output and applies the error difference from the output nodes back through the network to all the connections.

Back-propagation is an example of the \textit{delta learning rule} which aims to minimize the difference between actual and desired output using an \textit{error function}. If we imagine the error space as an n-dimensional surface \textit{(hypothesis space)}, the delta rule aims to minimize the error function by following the gradient of the surface through what is called \textit{gradient descent} \cite{bash}.

How much the weights are adjusted by is controlled by the learning rate, which dictates how quickly the network trains itself. If it is too fast, the ANN can come to early conclusions that may miss out other potentially better solutions in the problem space. If it is too low, the network may take too long to learn \cite{glor}.

Error functions are implemented through \textit{optimization algorithms}. Some commonly used optimization algorithms that will be used in this study are explained below \cite{optim}.\\

\textbf{SGD:} Stochastic gradient descent is traditional gradient except every iteration approximates the gradient based on a randomly picked sample from the data \cite{bot}.
  
\textbf{Rprop:} Resilient propagation is traditional back-propagation except the error propagated back is based on the sign of the error with a set update ($\Delta$) value which is modified on a batch basis \cite{mush}. A batch is just multiple data items sent to a neural network at once during processing \cite{hint}.
  
\textbf{RMSprop:} Root mean squared propagation is Rprop with SGD using smaller mini-batches. It scales the connection weight updates across the mini-batches by an exponentially decaying average of the SGD gradient squared \cite{mis}.
  
\textbf{Adam:} The Adam algorithm (adaptive moment estimation algorithm) builds from RMSprop by adding an exponential moving mean of the gradient and squared gradient. Two parameters then control the decay of these moving averages \cite{king2}.
  
\textbf{Adamax:} Adamax is a variation of the Adam algorithm in which the algorithm's parameters are influenced by fewer gradients, making it more resilient to noise \cite{king2}.

\subsection{Convolutional Neural Networks (CNNs)}
ANNs are good at recognizing patterns, which is what is needed to identify a galaxy from an image \cite{bash, amores}. These patterns can be seen as certain features that define a galaxy as outlined in \textit{section \ref{astro}} and any other patterns that the ANN might see that humans may not necessarily pick up \cite{khalifa, kim}. A CNN exploits this by performing preprocessing on an image (or n-dimensional array of data) to extract these features more definitively \cite{kriz}.

The CNN architecture consists of a \textit{convolutional layer, pooling layer, rectifier linear unit (ReLU)} and a \textit{fully-connected layer} \cite{simon}. The input layer takes in pixels and their values. The convolutional layer breaks the image into regions (\textit{filters}) and computes the output for neurons in these local regions while the ReLU layer applies an activation function on the regions\textquotesingle\ outputs. The ReLU function is very simple as it is just; \textit{$f(x) = max(0, x)$}, and is quick to evaluate. Another reason it helps speed up processing times is that all zero values it produces for any negative values create sparser intermediate representations of the input data. The pooling layer then performs down sampling on the image (producing a \textit{feature map}) before the fully connected layer (which is just an ANN) classifies the image. The convolutional, ReLU, and pooling layers can be repeated any number of times before the fully-connected layer is used \cite{simon, kriz}.

Both Kim et al. (2017) \cite{kim}, and Bottou (1991) \cite{bot} used CNNs in classifying astronomical objects with much success. CNNs can improve processing times since pooling together groups of features or pixels decreases the resolution and number of nodes needed in the fully-connected layer \cite{zhang}.

\subsection{Evolutionary Algorithms (EAs)}
EAs are based on the principle of Darwinian biological evolution. A set of potential solutions (population) to a problem or task represents a pool of individuals in a single generation. These individuals compete with each other to be the best at solving a problem or completing a task. How well an individual does is measured by its \textit{fitness} which is a function written by the coder that takes one or more factors into consideration to determine how well that individual is doing \cite{streich}. Once the fitness of every individual has been calculated, the fittest are then selected for the next generation before a technique called \textit{cross-over} is used. Cross-over combines the parameters or data from two or more fit individuals (parents) to breed more ``children'' to make up the population size for the next generation. Cross-over is a \textit{genetic} operator as it mimics the way genes from parent animals or humans are combined in their children \cite{holland}. \textit{Mutation} can also be programed in which randomly changes the parameters or data defining an individual in each generation \cite{streich}.

The process is repeated across multiple generations until it is stopped or reaches some optima, by which time the individuals in the generation should be much better at solving the problem or completing the task at hand than the randomly initiated starting population. If the problem space can be represented in an \textit{$n$}-dimensional space (\textit{n} $>$ 1), then it is essential that there are enough individuals with enough variation between them to cover a large enough area of the problem space so that the algorithm can converge on what is hopefully a global optimum \cite{streich}. Sometimes the algorithm can get stuck on some smaller local optima. This issue can be mitigated by slowing the algorithm's convergence through increasing the population size, increasing the mutation rate, and randomly selecting some unfit individuals from each generation to make it to the next generation \cite{holland}.

EAs can be applied to ANNs as a method of learning through \textit{NEAT} (Neuro-Evolution of Augmenting Topologies), or \textit{HyperNEAT} (Hypercube-based NEAT) \cite{low}, where the weights of node connections are evolved. An extension of this is DeepNEAT where the topology and hyperparameters of the ANN are evolved \cite{miik}. A \textit{hyperparameter} is a \textit{parameter} that is not learned within a network, while a parameter is the more broad term. Throughout this paper \textit{parameter} will be used as a general term.

\section{METHODS} \label{methods}
\subsection{Data Handling} \label{data}
Since the VISTA and UKIDSS surveys contain newer images that have a good quality and resolution \cite{saito, lucas}, they were used as the source for all training and testing data. UKIDSS and VISTA also provided the only data that is relatively well labelled in the ZoA with regards to galaxies and non-galaxies.

The survey data is presented in multiple sets of three FITS (Flexible Image Transport System) files, with each file in a set representing a different passband \cite{ponz}. FITS is a digital file format for storing n-dimensional array data (such as images) that may include multiple headers and is specifically tailored for scientific use \cite{han}. In the data there were files that labelled the given FITS files as galaxies and non-galaxies. The FITS images themselves ranged from large images containing multiple references to different objects in the image, to small images focused on one object. Overall the data amounted to 34GB.

The program (Galyxi Vysion) extracts the data into four 2D arrays corresponding to the different passbands of the image and its labelling data. Using this labelling information, the image is passed into a \textit{bounding-box} algorithm which creates sub-images by reducing the image onto one or more points in that image so that the sub-images' boundaries match the edges of a labelled object. The bounding-box algorithm works by incrementally decreasing the image size onto the focused point while constantly monitoring the next image's change in pixel-value mean and standard deviation. When there is a large enough change (decided by the experimenter), the image is considered reduced onto the labelled object.

Galyxi Vysion then normalize the images. Normalization also helps extract information about low surface brightness diffuse galaxies and galaxies obscured behind brighter stars in front of them \cite{butler, abra, jarreta, jarretb}. Each image is normalized based on its own pixel-value mean and standard deviation (as this produced the best results) before all the images are finally resized to the same size.

In the data acquired there were 3265 J-band images, 2559 H-band images, and 3403-K band images that could be extracted from the FITS images. The mismatch in the number of J, H, and K images is due to the fact that the data was not perfectly organized and contained some unmatched images along with duplicate images that had to be removed. Since the number of J, H, and K images were not equal, the number of JHK images that could be created was limited to 2505. The data was also not ideal and contained a lot of noise, such as images with inconsistencies between their respective J, H, and K counterpart images. These inconsistencies included; not being aligned or rotated correctly, not having the same resolutions or zooms, having off-center labelling, and glitches towards the edges of a few images. A \textit{data\_manager} program was written to help organize and match the different passband images while removing duplicates. Although the most inconsistent data was removed with the help of the \textit{data\_manager} program, the data still contained some imperfections.

  Of the 2505 JHK images found, 1278 were labelled as galaxies and 1232 were labelled as non-galaxies. Since this was all the raw data that could be acquired in the given time-frame, and more is usually needed to effectively train a CNN, some transformations were applied to the data in order to bulk its contents \cite{schmid, kriz, fayad}. These transformations included rotating the images by 90, 180, and 270 degrees in combination with reflections about the \textit{y}-axis.

For each transformation, different focus depths in the \textit{bounding-box} algorithm were also applied (if possible for the given image). This increased the number of images from 2505 to 22752, including 11808 galaxies and 10944 non-galaxies.

The labelling data classifies objects on a scale from definite galaxies, to probable galaxies, and finally definite non-galaxies. The intention of the study was to create a program that is able to automatically detect and identify galaxies in the ZoA to help astronomers provide a \textit{full} coverage of the sky \cite{kraan}. It is therefore less of an issue if the program makes false positives as compared to false negatives. This is because a false positive can easily be checked by a human and dismissed while making catalogues. However, in the case of a false negative, the location of the missed galaxy is still unknown. It is for this reason that probable galaxies were all seen as galaxies by Galyxi Vysion.

\subsection{CNN Implementation}
The first step in developing Galyxi Vysion was to create a working CNN class which was designed in a three-dimensional structure to best match the format of the JHK passbands. The data for an image is passed into the CNN with each pixel being allocated its own input node. The \textit{$x$} and \textit{$y$} axes of the input nodes correspond to the \textit{$x$} and \textit{$y$} coordinates of the input pixel while the different passbands correspond to the \textit{$z$} axis of the input nodes.

The input nodes feed into the first convolutional layer. All convolutional layers are followed by a ReLU layer before the output is passed to a pooling layer. The CNN class is set up to accept several parameters from the EA. The combination of parameters define the CNN's \textit{chromosome} with regards to the genetic approach of the EA \cite{holland}. The cycle from convolutional to pooling layer is repeated one or more times before the output is passed on to the fully connected layer which uses a binary classification of \textit{galaxy} or \textit{non-galaxy}.

The parameters that can be modified by the EA can be seen in \textit{table \ref{tests}}.

The fully-connected layer's activation function can be modified by the EA to either; ReLU, Leaky ReLU, ELU, Sigmoid, TanH, and Softplus, as outlined in \textit{section \ref{activations}}. These were chosen since they are commonly used activation functions \cite{shar} and are hence likely to achieve good results. The activation function parameter is an array that assigns a value to each of the above mentioned types and it is these values that are mutated and crossed. The highest value corresponding to an activation type in the parameter array is what is chosen to be used by all nodes in the fully-connected layer of a particular CNN.

In order to make the CNNs learn, the algorithms outlined in \textit{section \ref{optimizations}} (Rprop, SGD, RMSprop, Adamax and Adam) were used as possible optimization algorithms that the EA could select for a CNN individual to use in its fully-connected layer. These algorithms were also selected as they are commonly used in CNN implementations \cite{optim} and are hence likely to achieve good results. The optimization algorithm parameter is also an array that functions similarly to the activation function parameter.

\subsection{EA Implementation}
The method of this study is to use an EA to evolve the topology of a CNN \textit{(see section \ref{methods})}, which is similar to a DeepNEAT implementation. However, due to the extremely long processing times associated with NEAT \cite{miik}, and known success of non-evolutionary based gradient descent methods (as outlined in \textit{section \ref{optimizations}} \cite{schmid}, an evolutionary approach to topological configuration (as in DeepNEAT) was chosen without a neuroevolutionary approach to each individual CNN in the population.

The EA developed was designed to evolve the topology and parameters of a population of CNNs to find the best configuration for identifying and classifying galaxies in the ZoA. Galyxi Vysion works by taking a population of \textit{x} CNNs and having them all train on a random subset of the input data. After all the CNNs are trained, they are tested on a common set of testing data, using the percentage accuracy obtained by each CNN as their fitness. The program also records each CNNs training time, but it does not form a part of the fitness function. The program selects the fittest (50\%) of the CNNs to make it to the next generation where they are randomly paired up and have their parameters combined using cross-over to breed enough children to make up the rest of the population for the next generation.

In order to mitigate getting stuck in local optima, a program setting allows a small percentage of unfit CNNs to be randomly selected to become parents of the next generation. After parent CNNs have finished creating children, they are also subject to mutation which randomly alters some of the CNN's parameters (which are outlined in \textit{table \ref{tests}}).

The EA itself has parameters that define how it is run which can also be seen in \textit{table \ref{tests}}.

A set percentage of the data is randomly allocated by Galyxi Vysion to be training data while the rest is used for testing. The data was split exactly 50:50 for testing and training. Once the EA has run for all generations as specified in \textit{table \ref{tests}}, the program identifies the fittest CNN configuration it can find and uses all 100\% of the input data to train the CNN. The trained CNN is then saved so it may be used later on unseen data to potentially discover new galaxies.

\subsection{Program Development}
\subsubsection{Software}
Since Galyxi Vysion makes use of an EA to evolve CNNs, the computation time required to allow the program to learn was very long and with limited time for the project, a way of maximizing the efficiency of the EA and CNNs was needed.

CNNs make use of neural networks which are easily parallelizable \cite{simon}. Pytorch is a machine learning library for Python which allows easier parallel implementation of neural networks along with support for CUDA, and is hence what was used to code Galyxi Vysion.

CUDA allows a program to exploit the massively parallel abilities of a graphics card to process data. The batch size used had a large impact on performance, with a larger batch size allowing more data to be fed in parallel to the GPU. On the other hand, a larger batch size means slower convergence. A balance of 10\% of the overall number of input images was decided upon as the batch size when training CNNs in the EA. For 22752 JHK input images the batch size was therefore $ceil(22752 / 10) = 2276$.

\subsubsection{Hardware} \label{hardware}
A difficulty with CUDA was being able to keep the GPU busy as the CPU struggled to feed data to the GPU fast enough to use 100\% of the GPU. All tests were run on a single machine using an Nvidea 1080 Ti, 6 CPU cores at 4.6GHz, and CUDA 9.2. In order to get the maximum usage out of the GPU, Galyxi Vysion is able to move all the data to the GPU's VRAM before starting the evolutionary process. This ensures that there is no bottleneck from the rest of the system as moving CUDA tensors into GPU memory takes more time than processing them \cite{lux}.

\subsubsection{Program Structure}
All the settings for the program including the parameters for the EA and CNN classes can be modified via an input file. The program also displays and outputs to file the details regarding the EA's progress so that the data can be collected once a test is complete. As the code runs, the progress of the EA is saved at the start of each generation. This means that if the software or hardware crashes for whatever reason in the middle of a test, not all progress would be lost. It also enables a test to be stopped and moved to a different machine. Two more subprograms were written to help organize the raw data \textit{(see section \ref{data})} and to generate statistical graphs on the details of a test once complete. A final subprogram named \textit{Galyxi Fynder} was created to use a trained CNN on unseen data and potentially discover new galaxies.

\section{EXPERIMENTS AND EVALUATION} \label{experiments}
\subsection{Experimental Design}

Before testing was done using data from the ZoA, small scale tests were done using the popular CIFAR10 dataset to see if everything was running as it should and that the CNNs both learned, tested, and evolved correctly. The J, H, and K channels in the ZoA images substitute the RGB channels in the CIFAR10 images.

All tests were used to gather information on how well the EA worked on the ZoA data, how fast convergence took place, and how the CNNs evolved under different EA parameters and program settings. The results from these tests were used to configure the settings used for a \textit{final test} to see if better results could be obtained with human parameter tuning. In these tests, the settings that control the maximum and minimum values that can be evolved by a CNN in the EA were set as wide as was logical with regards to the dimensions of the input data and expected processing time. There were also further limitations on these parameters that were put in place in order to avoid errors for invalid CNN configurations that the EA might produce.

Since the data that could be obtained in the given time-frame for this study was noisy, and a lot of the noise was due to inconsistencies between the J, H, and K passband images, tests were also done using J, H, and K images separately to see if different results would be obtained. After transformations, 27496 J, 21304 H, and 30080 K images were created. Testing individual bands also validates the necessity of having to use all J, H, and K passbands together.

Before Galyxi Vysion resizes the input images while data-handling \textit{(see section \ref{data})}, the average input image size of the created JHK images was 126x126 while the average J, H, and K image sizes were 111x111, 125x125, and 124x124 respectively. Tests were done at four different image resize values to see if using larger images obtained better results, since smaller images are faster to process. The image size used in the CIFAR10 dataset is only 32x32, even though the images it contains are quite complex with 10 possible classifications. This prompted the first image input size tested to be 32x32. After 32x32 was tested, 64x64 was used before finally testing at 24x24 and 16x16 with separate tests done on JHK, J, H, and K images in each case. The reason dimensions as small as 16x16 and 24x24 were used was that the results obtained at 64x64 did not provide good enough results to warrant testing at larger sizes.

\subsection{Evaluation}

Once all tests (excluding the final test) were done, the results were evaluated and used to select the most optimal input image size (with regards to processing time and results), EA settings, and EA boundary parameters for the final test. This was done by evaluating graphs generated by the tests which show the spread of parameter values chosen by the EA for a given test across all generations run against the average fitness seen by individual CNNs using those parameter values. With this, boundary values for the final test could be adjusted so as to discard parameter values that were consistently bad or not used, or to add new values that may need to be included in the problem space. Graphs depicting the evolutionary convergence as the average fitness per generation for each test were also generated and were used to adjust the EA settings for the final test.

From the graphs produced, the gradients of different locations in the problem space along with various optima can be seen for individual parameters. This is especially easy to see in the the number of fully-connected features (nodes in a layer) against average fitness. If the plot gravitates around one or more optimal parameter values, then outlying values that are largely unsuccessful can be discarded when parameterizing the final test. Also, if the gradient of a parameter's value is increasing to the edge of the bounds, then these bounds can be widened.

Once the settings and parameters for the final test had been determined from the results of the other tests, the final test was run and its results were compared to the other test results. The non-final test results were also evaluated statistically using t-tests to see if there is a significant difference between image input sizes and passbands used.

A table of the input boundary parameters given to the EA during all tests can be seen in \textit{table \ref{tests}}. The maximum boundary values for the 64x64 input image tests are separate from the maximum boundary values for the 32x32, 24x24, and 16x16 input image tests. This is because the number of input nodes corresponding to pixels in the 64x64 image tests requires a larger network to sufficiently cover the problem space. For images smaller than 32x32, the problem space is small enough that it is inconsequential to keep the same boundary values as the 32x32 image boundaries. The EA settings used for all tests are displayed in \textit{table \ref{evo}}.

Note that the learning rate is as a learn-able parameter, implemented similarly to the study done by Miikkulaine et al \cite{miik}. For clarification, an \textit{epoch} is one forward and backward pass of all training examples \cite{ann}.

\begin{table}[t]
\fontsize{6.5}{8}\selectfont
  \caption{EA settings for all tests.}
  \label{evo}
  \begin{tabular}{|m{5cm}|m{1cm}|m{1cm}|}
    \hline \centering  \textbf{Setting} & \centering \textbf{Other Tests} & \centering \textbf{Final Test} \tabularnewline
    \hline \centering  number of generations & \centering 50 & \centering 20 \tabularnewline
    \hline \centering  population size & \centering 100 & \centering 100 \tabularnewline
    \hline \centering  epochs each CNN learns for in a generation & \centering 10 & \centering 10 \tabularnewline
    \hline \centering  percentage of fit CNNs allowed to breed & \centering 50\% & \centering 50\% \tabularnewline
    \hline \centering  highest percentage of unfit CNNs randomly allowed to breed & \centering 2\% & \centering 5\% \tabularnewline
    \hline \centering  chance that a CNN will be mutated & \centering 0.05 & \centering 0.10 \tabularnewline
    \hline \centering  chance that a parameter will be mutated in a CNN selected for mutation & \centering 0.10 & \centering 0.20 \tabularnewline
    \hline
  \end{tabular}
\vspace{-2mm}
\end{table}

\begin{table}[t]
\fontsize{6.5}{8}\selectfont
  \caption{EA parameter boundaries for all tests.}
  \label{tests}
  \begin{tabular}{|m{3cm}|m{0.7cm}|m{0.7cm}|m{0.7cm}|m{1.3cm}|}
    \hline \centering  \textbf{Setting} & \centering \textbf{Min All} & \centering \textbf{Max 32x32} & \centering \textbf{Max 64x64} & \centering \textbf{Max Final Test} \tabularnewline
    \hline \centering  convolutional and pooling layers & \centering 1 & \centering 4 & \centering 8 & \centering 2 \tabularnewline
    \hline \centering  fully-connected layers & \centering 1 & \centering 8 & \centering 16 & \centering 4 \tabularnewline
    \hline \centering  convolutional output channels & \centering 3 & \centering 36 & \centering 72 & \centering 30 \tabularnewline
    \hline \centering  convolutional kernel size & \centering 2 & \centering 8 & \centering 16 & \centering 16 \tabularnewline
    \hline \centering  convolutional kernel stride & \centering 1 & \centering 4 & \centering 8 & \centering 4 \tabularnewline
    \hline \centering  convolutional padding & \centering 0 & \centering 3 & \centering 6 & \centering 6 \tabularnewline
    \hline \centering  convolutional groups & \centering 1 & \centering 3 & \centering 3 & \centering 1 \tabularnewline
    \hline \centering  pooling kernel size & \centering 2 & \centering 8 & \centering 16 & \centering 16 \tabularnewline
    \hline \centering  pooling kernel stride & \centering 2 & \centering 4 & \centering 8 & \centering 8 \tabularnewline
    \hline \centering  pooling padding & \centering 0 & \centering 3 & \centering 6 & \centering 6 \tabularnewline
    \hline \centering  fully-connected features (nodes) & \centering 3 & \centering 256 & \centering 512 & \centering 160 \tabularnewline
    \hline \centering  learning rate & \centering 0.0001 & \centering 0.01 & \centering 0.01 & \centering 0.004 \tabularnewline
    \hline \centering  optimization algorithms &
    \multicolumn{3}{c|}{all} & \centering Exclude SGD \tabularnewline
    \hline \centering  activation functions &
    \multicolumn{4}{c|}{all} \tabularnewline
    \hline
  \end{tabular}
\end{table}

\section{RESULTS AND DISCUSSION}
\subsection{Results}

\begin{figure*}[t]
\caption{Graphs of optimal test run with 32x32 JHK input images.}
  \subfloat{\includegraphics[width = 0.2\linewidth]{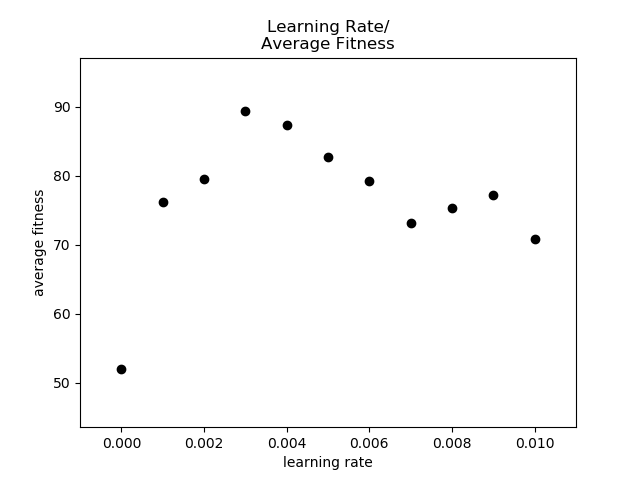}}
  \subfloat{\includegraphics[width = 0.2\linewidth]{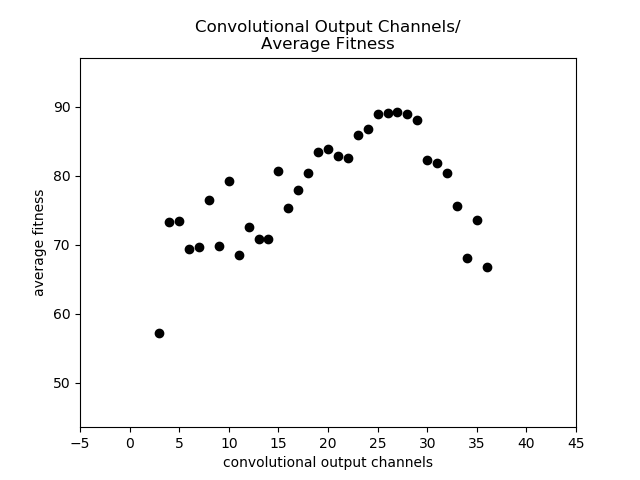}}
  \subfloat{\includegraphics[width = 0.2\linewidth]{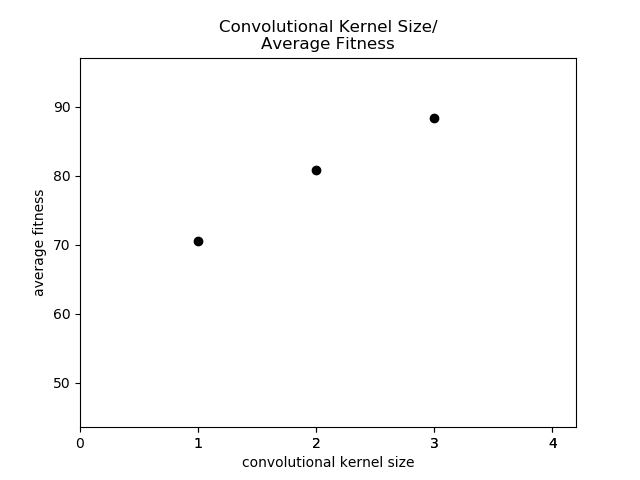}}
  \subfloat{\includegraphics[width = 0.2\linewidth]{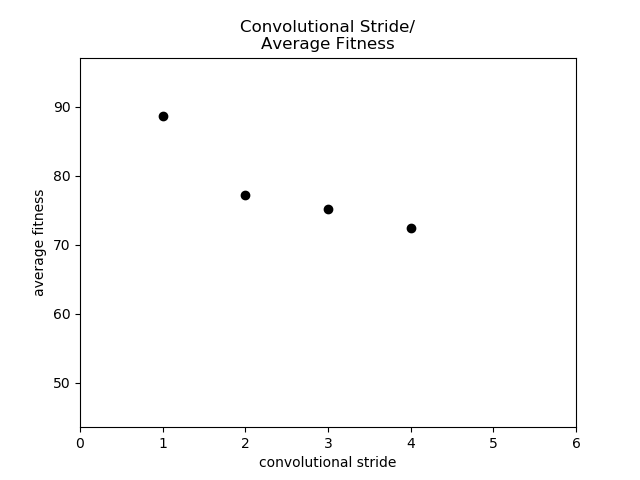}}
  \subfloat{\includegraphics[width = 0.2\linewidth]{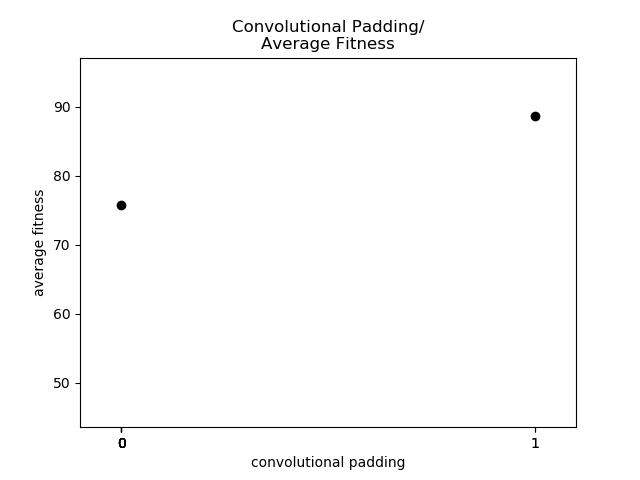}}\\
  \subfloat{\includegraphics[width = 0.2\linewidth]{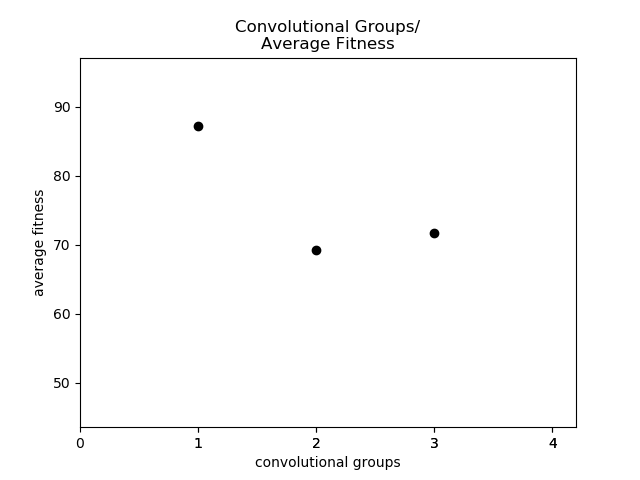}}
  \subfloat{\includegraphics[width = 0.2\linewidth]{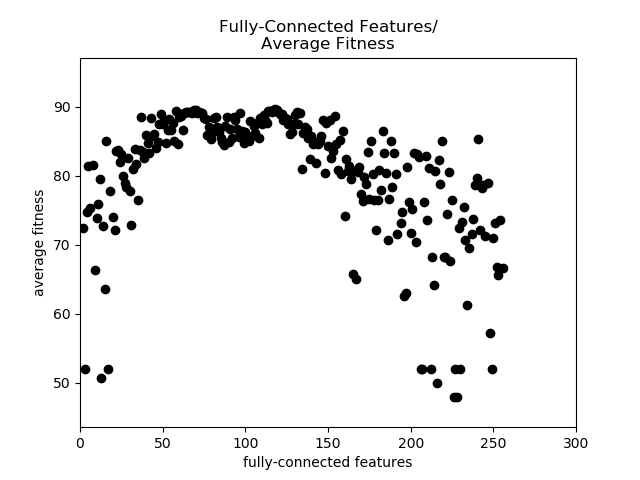}}
  \subfloat{\includegraphics[width = 0.2\linewidth]{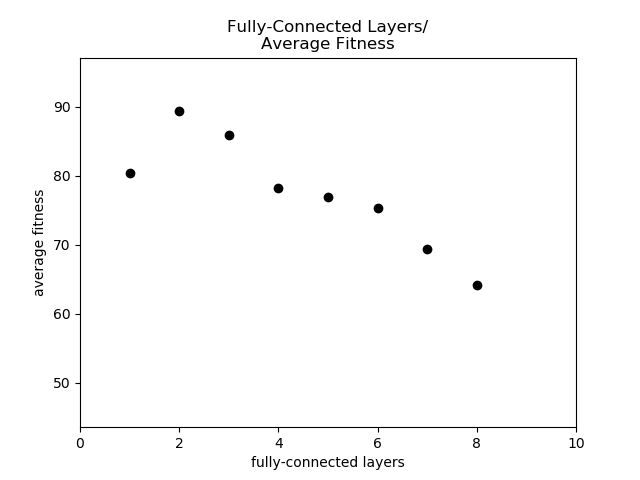}}
  \subfloat{\includegraphics[width = 0.2\linewidth]{learning_rate.png}}
  \subfloat{\includegraphics[width = 0.2\linewidth]{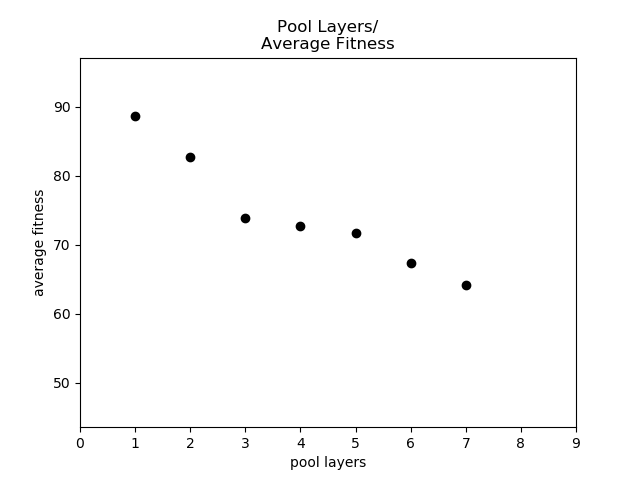}}\\
  \subfloat{\includegraphics[width = 0.2\linewidth]{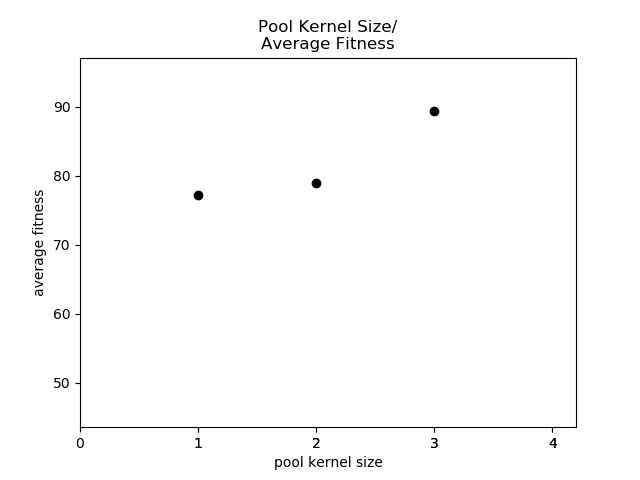}}
  \subfloat{\includegraphics[width = 0.2\linewidth]{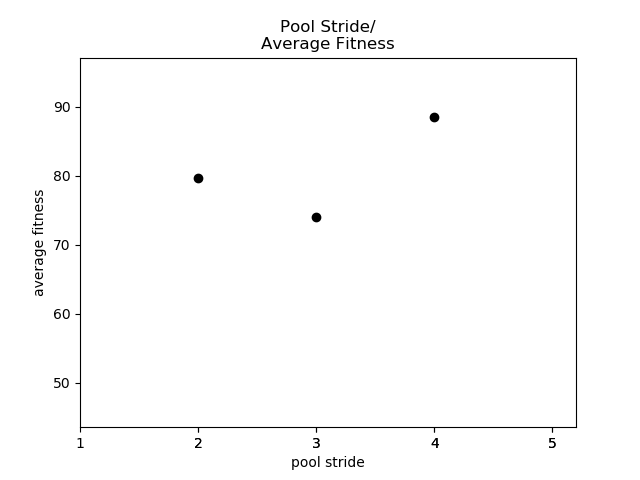}}
  \subfloat{\includegraphics[width = 0.2\linewidth]{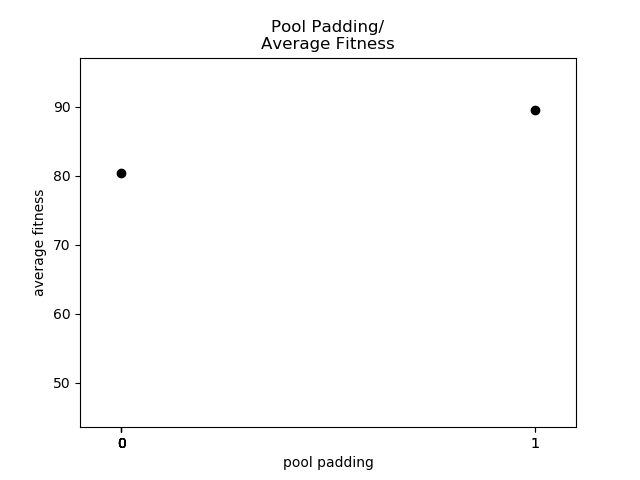}} 
  \subfloat{\includegraphics[width = 0.2\linewidth]{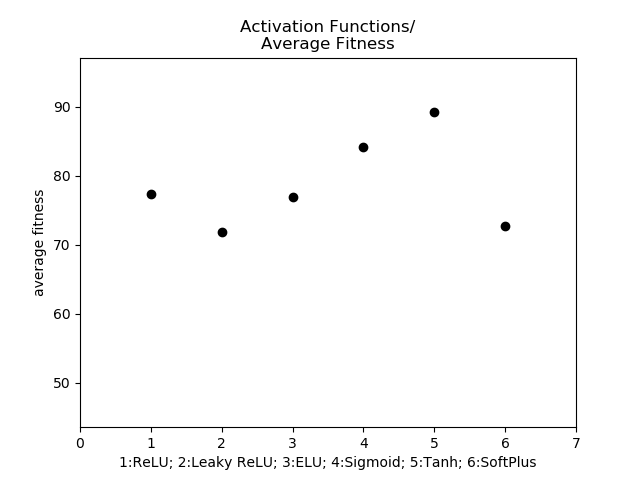}}
  \subfloat{\includegraphics[width = 0.2\linewidth]{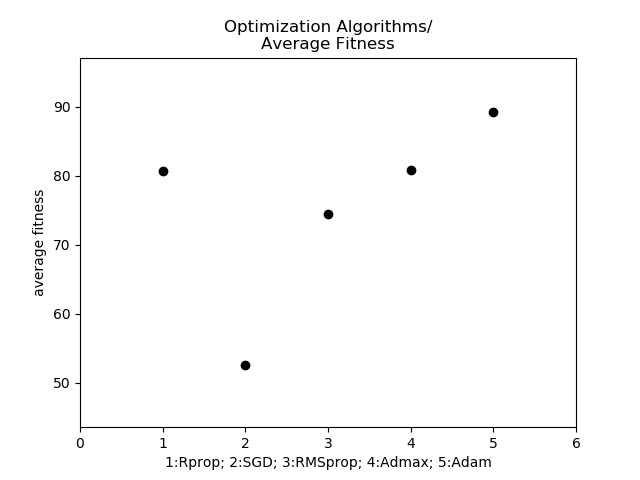}}
  \label{graphs}
\vspace{-2.5mm}
\end{figure*}

The best test result achieved a percentage accuracy of 91.75\% using an input image size of 32x32 with JHK images. The graphs generated by this test and the optimal CNN configuration produced can be seen in \textit{figures \ref{graphs}} and \textit{\ref{config}} respectively. The results regarding the best CNNs found across all tests excluding the final test can be seen in \textit{table \ref{results}}.

\begin{table}[t]
\fontsize{6.5}{8}\selectfont
  \caption{Best CNN \% accuracies for all image input sizes and passbands used with averages \textbf{$\mu$} and standard deviations \textbf{$\sigma$}.}
  \label{results}
  \begin{tabular}{|m{1.4cm}|m{0.8cm}|m{0.8cm}|m{0.8cm}|m{0.8cm}|m{0.6cm}|m{0.6cm}|}
    \hline \centering  \textbf{Passbands Used} & \centering \textbf{64x64 Input} & \centering \textbf{32x32 Input} & \centering \textbf{24x24 Input} & \centering \textbf{16x16 Input} & \centering \textbf{$\mu$} & \centering \textbf{$\sigma$} \tabularnewline
    \hline \centering JHK & \centering 88.92 & \centering 91.75 & \centering 90.61 & \centering 89.77 & \centering  90.26 & \centering 1.05 \tabularnewline
    \hline \centering J & \centering 87.14 & \centering 88.20 & \centering 87.73 & \centering 86.88 & \centering 87.49 & \centering 0.51 \tabularnewline
    \hline \centering H & \centering 85.49 & \centering 87.96 & \centering 87.66 & \centering 86.82 & \centering 86.98 & \centering 0.92 \tabularnewline
    \hline \centering K & \centering 89.12 & \centering 90.59 & \centering 88.88 & \centering 88.12 & \centering 89.18 & \centering 0.90 \tabularnewline
    \hline \centering \textit{$\mu$} & \centering 87.67 & \centering 89.63 & \centering 88.72 & \centering 87.90 & \centering - & \centering - \tabularnewline
    \hline \centering \textit{$\sigma$} & \centering 1.47 & \centering 1.58 & \centering 1.19 & \centering 1.20 & \centering - & \centering - \tabularnewline
    \hline
  \end{tabular}
\vspace{-2mm}
\end{table}

\begin{figure}[t]
\caption{Convergence for the optimal test using 32x32 JHK input images (left) and the final test (right).}
  \subfloat{\includegraphics[width = 0.44\linewidth]{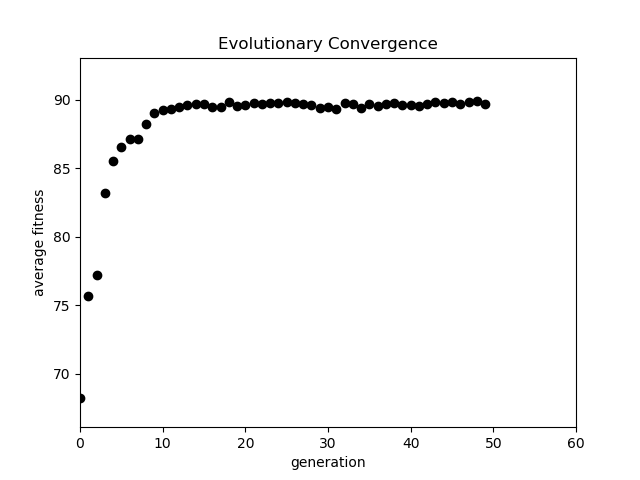}}\hspace{4mm}
  \subfloat{\includegraphics[width = 0.44\linewidth]{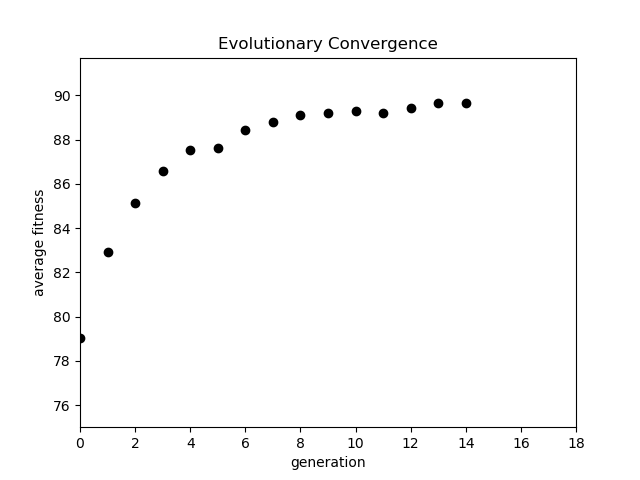}}
  \label{converge}
\vspace{-2mm}
\end{figure}

\begin{table}[t]
\caption{Best CNN configuration using 32x32 JHK images.}
\fontsize{6.5}{8}\selectfont
  \begin{tabular}{|m{1.8cm}|m{1cm}|m{1cm}|m{0.8cm}|m{0.8cm}|m{0.8cm}|}
    \hline \centering  \textbf{Layer} & \centering \textbf{Output Channels} & \centering \textbf{kernel Size} & \centering \textbf{Stride} & \centering \textbf{Padding} & \centering \textbf{Groups} \tabularnewline
    \hline \centering  convolutional 1 & \centering 25 & \centering 3 & \centering 1 & \centering 1 & \centering 1 \tabularnewline
    \hline \centering  pooling 1 & \centering n/a & \centering 3 & \centering 4 & \centering 1 & \centering n/a \tabularnewline
    \hline \centering  fully-connected 1 & \centering 25 & \centering n/a & \centering n/a & \centering n/a & \centering n/a \tabularnewline
    \hline \centering  fully-connected 2 & \centering 25 & \centering n/a & \centering n/a1 & \centering n/a & \centering n/a \tabularnewline
    \hline
  \end{tabular}
  \begin{tabular}{|m{1.4cm}|m{0.5cm}|m{1.2cm}|m{0.6cm}|m{1.6cm}|m{0.8cm}|}
    \hline \centering  \textbf{learning rate:} & \centering 0.003 & \centering \textbf{activation function:} & \centering TanH & \centering \textbf{optimization algorithm:} & \centering ADAM \tabularnewline
    \hline
  \end{tabular}
  \label{config}
\vspace{-2mm}
\end{table}

After performing independent samples t-tests on all pairs of tests done across the different input image sizes (as shown in \textit{appendix \ref{statsizes}}), the lowest p-value found was 0.17. Using the standard level of significance of 0.05 \cite{lac}, the lowest p-value is still greater than 0.05, indicating there is no significant difference across the different image input sizes. This means that the network is not getting any benefit from the extra detail in larger images, and is simply wasting processing power. Therefore, a large input image is not needed to identify galaxies in the ZoA. However, the best result was still obtained using an input size of 32x32, which is why 32x32 was chosen as the input size for the final test.

Using independent samples t-tests on all pairs of tests done using different passbands (as shown in \textit{appendix \ref{statpassbands}}) shows that using JHK images is significantly better than just using J or H images since the respective p-values of 0.006 and 0.007 are lower than 0.05 \cite{lac}. It also reveals that there is no significant difference in using either J or H, and that K images are also not significantly different from JHK Images. However, the JHK images also had the most noise as all inconsistencies between the respective J, H, and K parts of the JHK image in the dataset were relevant, and in all single band tests these inconsistencies were no longer an issue. Since the JHK images already performed the best on average, there is a potential that they could have done even better had there not been inconsistencies in the data. For this reason, JHK images were chosen to be used in the final test.

For the test using 32x32 JHK images, the EA's convergence graph (shown in \textit{figure \ref{converge}}) indicates that the algorithm converged in only just over 10 generations, but ran for 50 generations in 9 hours \textit{(see section \ref{hardware} for system hardware)}. This meant that for the final test the mutation rate could be increased in order to slow the convergence \cite{streich} and potentially find a better configuration, while the maximum number of generations was decreased.

By further analyzing the 32x32, JHK test, it can be seen that the most successful activation functions included \textit{Sigmoid} and \textit{TanH} while having one convolutional group worked best. The number of convolutional layers, convolutional stride and pooling layers achieved good results with smaller values, as seen in \textit{figure \ref{graphs}}.

The number of fully connected features appears to be bimodal around 70 and 125 while convolutional output channels is unimodal around 25 to 30. The best optimization algorithm was ADAM, with SGD performing poorly, and the learning rate seemed to get peek results between 0.002 and 0.004. It is also worth noting that different activation functions and optimization algorithms had better or worse results depending on the image input size and passbands used, which indicates how specific CNN parameter and topological configurations greatly impact results, hence highlighting the benefit of the EA. Only the SGD optimization algorithm was discarded for the final test.

\subsection{Discussion}

\subsubsection{Final Test Results}

The results from the final test produced a CNN configuration that achieved a percentage accuracy of 91.08\%, converged in under 15 generations, and took a much briefer time of two hours to process \textit{(see section \ref{hardware} for system hardware)}. The convergence graph can be seen in \textit{figure \ref{converge}}. This is, however, not the best result seen, which is still the first test at 32x32 using JHK images. This highlights the benefit of the EA algorithm, as by trying to parameterize the test to some extent manually, a better result could not be obtained.

\subsubsection{Analysis}

Identifying galaxies is a binary classification so by random chance a fitness (percentage accuracy) of 50 is expected if the data is given to an untrained network.

It is evident from the tests conducted that the image input size used had little effect on the results obtained for identifying galaxies in the ZoA. This could be due to a number of reasons. Firstly, the number of features that can be extracted from a light source such as a galaxy or star, is far less complex than the images in the CIFAR10 dataset, which can potentially be used to classify planes, cars, birds, cats, deer, dogs, frogs, horses, ships, and trucks from each other using only 32x32 images. This highlights a CNNs potential to extract features and classify them, and even with the noise from other light sources in the ZoA \cite{lucas, gonz, barav}, the features of galaxies can still be picked out at lower resolutions.

Another potential reason for this is that the data may have been biased in its examples of galaxies and non-galaxies. If the data consistently shows examples of non-galaxies as either being bright stars or empty space, then once the image has been focused on the labelled object, the CNN would only need to know if the brightness is within a certain range, and this could technically be given in one pixel.

The challenges associated with identifying galaxies in the ZoA include finding galaxies obscured by stars, finding low diffuse galaxies, and differentiating galaxies from similar looking objects such as planetary nebula and multi-star systems \cite{butler, abra, jarreta, jarretb}. If the labelling data does not provide enough of these examples, the CNN may consistently get the same examples wrong and the obvious examples right. This may also explain why regardless of the method used to run a test, the best result seems to be capped at around 90\% accuracy.

One way to test this would have been to have the labelled data divided up, with more classifications. These would include a classification for each galaxy and non-galaxy case, as mentioned above, with a large, and equal number of examples for each classification. This is particularly needed in separating low surface brightness diffuse galaxies from other galaxies, as they have very different characteristics. That way one could be sure that the system is able to learn how to identify different objects, including galaxies, within the ZoA. However, due to the restraints and limitations of the data obtained for this study, this was not possible.

One potential limitation of the Galyxi Vysion system is that the EA does not evolve activation functions on a per-layer basis, but rather applies a single activation function to all the fully-connected layers. Previously, success has been seen in networks that utilize different activation functions for different layers \cite{schmid}, and had this been implemented, better results may have ensued.

The fact the that final test did not perform as well as the best of the other tests shows that the best results are most likely to be obtained by widening the search space and letting the EA find the optimal configuration. This prompts the suggestion that even more parameters with wider ranges could have been modifiable by Galyxi Vysion with a larger population, however this was out of the scope of the project and may not have produced better results due to the data available.

Once all tests were done the optimal CNN was applied using Galyxi Fynder to a FITS image of 20690 cataloged light sources, from which it was able to identify 3907 potential galaxies.

\section{CONCLUSIONS}

In support of the hypotheses, the implementation of an EA to evolve the topologies of a population of CNNs was successful as it was able to produce a unique CNN configuration that could identify galaxies in the ZoA with an accuracy of 91.75\% using only 32x32 images. With this accuracy level, the optimal CNN was able to generalize about the characteristics of galaxies through the noise of the ZoA, and could be applied to unseen data to identify the potential locations of galaxies in the ZoA.

The final test converged in under 15 generations with a relatively short processing time, given the hardware available. In agreement with the hypothesis made, this makes the evolutionary approach worthwhile over a human adjusted CNN configuration. A human is still needed to verify what has been flagged. However, should better data be available in the future, Galyxi Vysion could be trained again to evolve a new optimal CNN that may produce better and more reliable results.

Outside the ZoA, Odenwahn et al \cite{oden} used Neural Networks to identify galaxies, with accuracy results as high as 95\% - 99\% \cite{drink}. Both Khalifa et al \cite{khalifa} and Kim et al \cite{kim} could successfully classify galaxies outside the ZoA, with Khalifa et al achieving a 97\% accuracy. However, the attempt by Lahav et al \cite{lahav} near the ZoA only achieved 80\% - 96\%. These attempts begin to become unviable directly in the ZoA, such as the attempt by Drinkwater et al \cite{drink}. As hypothesized, when compared to these results, the  91.75\% achieved in this study directly in the ZoA is very successful, considering the difficulty in obtaining good data in the ZoA, which is one of the main reasons why other attempts have avoided it \cite{khalifa, kim, kriz}.

The fact that 3907 possible galaxies were identified in unseen data using Galyxi Fynder highlights the potential of Galyxi Vysion as a tool for creating full sky catalogs of galaxies \cite{lucas, kraan}. This could be used to help astronomers in many fields of research such as the \textit{dipole} in the \textit{Cosmic Microwave Background} (CMB), peculiar velocity of galaxies in our local cluster \cite{giov}, isotropy of our local universe and understanding the \textit{Great Attractor} \cite{ray}.

As predicted in the hypotheses, using all J, H, and K passbands provided better results, even if marginally.  However, in contradiction to what was hypothesized, the study shows that very little information in terms of input resolution is needed to identify a galaxy. This is in fact promising, as learning on small images is a lot faster, allowing for more parameters and wider parameter boundaries to potentially be used to find better configurations in the future.

\newpage

\bibliographystyle{ACM-Reference-Format}

\newpage

\begin{appendices}

\captionsetup{labelformat=empty}

\section{Activation Functions} \label{appendactive}

\begin{figure}[h]
  \tiny
  \begin{tabular}{|m{0.4cm}|m{0.7cm}|m{2.1cm}!{\vrule width 1.2pt} m{0.5cm}|m{0.7cm}|m{1.7cm}|}
    \hline \centering \textbf{Name} & \centering \textbf{Graph} & \centering \textbf{Function} & \centering \textbf{Name} & \centering \textbf{Graph} & \centering \textbf{Function} \tabularnewline
    \hline \centering ReLU & \centering
      \begin{tikzpicture}[scale=0.3, declare function={func(\x)= (\x < 0) * (0) + and (x >= 0,\x >= 0) * (\x);}]
        \begin{axis}[
        axis x line=middle, axis y line=middle,
        ymin=-3, ymax=3, ytick={-3,...,3},
        xmin=-3, xmax=3, xtick={-3,...,3},
        ymajorgrids=true,
        xmajorgrids=true]
        \addplot[color=black, ultra thick]{func(x)};
        \end{axis}
        \end{tikzpicture} &
        \begin{flushleft}\(\textit{f(x) = }\begin{cases} 0 & x< 0 \\ x & x \geq 0 \\\end{cases}\)\end{flushleft} &
    \centering Leaky ReLU & \centering
      \begin{tikzpicture}[scale=0.3, declare function={func(\x)= (\x < 0) * (\x*0.15) + and (x >= 0,\x >= 0) * (\x);}]
        \begin{axis}[
        axis x line=middle, axis y line=middle,
        ymin=-3, ymax=3, ytick={-3,...,3},
        xmin=-3, xmax=3, xtick={-3,...,3},
        ymajorgrids=true,
        xmajorgrids=true]
        \addplot[color=black, ultra thick]{func(x)};
        \end{axis}
        \end{tikzpicture} &
        \begin{flushleft}\(\textit{f(x) = }\begin{cases} 0.01x & x< 0 \\ x & x \geq 0 \\\end{cases}\)\end{flushleft} \tabularnewline
    \hline \centering ELU & \centering
      \begin{tikzpicture}[scale=0.3, declare function={func(\x)= (\x < 0) * (e^\x - 1) + and (x >= 0,\x >= 0) * (\x);}]
        \begin{axis}[
        axis x line=middle, axis y line=middle,
        ymin=-3, ymax=3, ytick={-3,...,3},
        xmin=-3, xmax=3, xtick={-3,...,3},
        ymajorgrids=true,
        xmajorgrids=true]
        \addplot[color=black, ultra thick]{func(x)};
        \end{axis}
        \end{tikzpicture} &
        \begin{flushleft}\(\textit{f(x) = }\begin{cases} \alpha (e^{x} - 1) & x< 0 \\ x & x \geq 0 \\\end{cases}\)\end{flushleft} &
    \centering Sigmoid & \centering
      \begin{tikzpicture}[scale=0.3]
        \begin{axis}[
        axis x line=middle, axis y line=middle,
        ymin=-3, ymax=3, ytick={-3,...,3},
        xmin=-3, xmax=3, xtick={-3,...,3},
        ymajorgrids=true,
        xmajorgrids=true]
        \addplot[color=black, ultra thick]{1/(1+e^(-x))};
        \end{axis}
        \end{tikzpicture} &
        \begin{flushleft}\(\begin{aligned} \ \textit{f(x) = } \frac{1}{(1+e^{-x})} \end{aligned}\)\end{flushleft} \tabularnewline
    \hline \centering TanH & \centering
      \begin{tikzpicture}[scale=0.3]
        \begin{axis}[
        axis x line=middle, axis y line=middle,
        ymin=-3, ymax=3, ytick={-3,...,3},
        xmin=-3, xmax=3, xtick={-3,...,3},
        ymajorgrids=true,
        xmajorgrids=true]
      \addplot[color=black, ultra thick]{(2/(1+e^(-2*x))) - 1};
        \end{axis}
        \end{tikzpicture} &
        \(\textit{f(x) = } \frac{2}{(1+e^{-2x})} - 1\) &
    \centering SoftPlus & \centering
      \begin{tikzpicture}[scale=0.3]
        \begin{axis}[
        axis x line=middle, axis y line=middle,
        ymin=-3, ymax=3, ytick={-3,...,3},
        xmin=-3, xmax=3, xtick={-3,...,3},
        ymajorgrids=true,
        xmajorgrids=true]
      \addplot[color=black, ultra thick]{ln(1 + e^(x)))};
        \end{axis}
        \end{tikzpicture} &
        \(\textit{f(x) = } \log _{e} (1+e^{x})\) \tabularnewline
    \hline
  \end{tabular}
  \caption{Appendix A: Commonly used activation functions \cite{shar}.}
  \label{functions}
\end{figure}

\section{T-Tests}

\subsection{Image Size T-Tests} \label{statsizes}

\FloatBarrier
\begin{table}[h]
\fontsize{6.5}{8}\selectfont
  \caption{Appendix B.1: Independent samples t-tests and their respective p-values between all input image size pairs.}
  \begin{tabular}{|m{1.6cm}|m{1.6cm}|m{1.6cm}|m{1.6cm}|}
    \hline \centering \textit{Input Size} & \centering \textbf{32x32} & \centering \textbf{24x24} & \centering \textbf{16x16} \tabularnewline
    \hline \centering \textbf{64x64} & \centering 0.17 & \centering 0.37 & \centering 0.84 \tabularnewline
    \hline \centering \textbf{32x32} & \centering - & \centering 0.46 & \centering 0.18 \tabularnewline
    \hline \centering \textbf{24x24} & \centering - & \centering - & \centering 0.43 \tabularnewline
    \hline
  \end{tabular}
\end{table}
\FloatBarrier

\subsection{Passband T-Tests} \label{statpassbands}

\FloatBarrier
\begin{table}[h]
\fontsize{6.5}{8}\selectfont
  \caption{Appendix B.2: Independent samples t-tests and their respective p-values between all input image passband pairs.}
  \begin{tabular}{|m{1.6cm}|m{1.6cm}|m{1.6cm}|m{1.6cm}|}
    \hline \centering \textit{Passbands} & \centering \textbf{H} & \centering \textbf{K} & \centering \textbf{JHK} \tabularnewline
    \hline \centering \textbf{J} & \centering 0.452 & \centering 0.030 & \centering 0.006 \tabularnewline
    \hline \centering \textbf{H} & \centering - & \centering 0.027 & \centering 0.007 \tabularnewline
    \hline \centering \textbf{K} & \centering - & \centering - & \centering 0.221 \tabularnewline
    \hline
  \end{tabular}
\end{table}
\FloatBarrier

\end{appendices}

\end{document}